\documentclass[twocolumn,prb,floatfix]{revtex4}

\usepackage{graphicx}
\usepackage[usenames]{color}
\usepackage{subeqn}

\date{\today}

\begin{document}
%
%Newcommands:
\newcommand{\kpar}{k_\parallel}
\newcommand{\lint}[3]{\int\limits_{#1}^{#2}d{#3}\,}

%\color commands:
\newcommand{\cred}[1]{{\color{BrickRed}#1}}
\newcommand{\cblue}[1]{{\color{Blue}#1}}
\newcommand{\cgreen}[1]{{\color{PineGreen}#1}}
\newcommand{\corange}[1]{{\color{Orange}#1}}
\newcommand{\cbrown}[1]{{\color{Brown}#1}}

\title{Optimal operation of transition-edge sensors on ballistic membranes}
\author{T. K\"uhn}
%\affiliation{Nanoscience Center, Department of Physics, University of Jyv\"askyl\"a, P.O. Box 35, FIN-40014 University of %%@
%Jyv\"askyl\"a}
%
\author{I. J. Maasilta}

\affiliation{Nanoscience Center, Department of Physics, University of Jyv\"askyl\"a, P.O. Box 35, FIN-40014 University of %%@
Jyv\"askyl\"a, Finland}
\begin{abstract}\noindent
We calculate the operating parameters of a transition edge sensor that is mounted on
a thin dielectric membrane with the assumption that the phononic heat transport in
the membrane is ballistic. Our treatment uses the correct phonon modes from elasticity theory (Lamb-modes), and spans the %%@
transition from 3D to 2D behavior. The phonon cooling power and conductance have a global minimum
as function of membrane thickness, which leads to an optimal value for the membrane thickness with
respect to noise equivalent power at a fixed operating temperature. The energy resolution of a calorimeter will not be affected %%@
strongly, but, somewhat counterintuitively, the effective time constant can be reduced by decreasing the membrane thickness in the %%@
2D limit. 
\end{abstract}

%\pacs{}

\maketitle

%\section{Introduction}

Superconducting transition edge sensors (TES) are currently under heavy development to be used as ultrasensitive calorimeters and %%@
bolometers over a wide range of frequencies, from sub-mm radiation to gamma rays.\cite{enss} Some recent examples of ambitious %%@
projects using TES sensors are: the new sub-mm camera for the James Clerk Maxwell telescope (SCUBA-II) \cite{scuba}, the detection %%@
of single near-infrared photons for quantum cryptography \cite{ros}, the X-ray imaging spectrometers for future ESA, NASA and JAXA %%@
missions \cite{xrayappl}, and the $\gamma$-ray detection of nuclear materials \cite{zink}.  
In most of these detector designs, the superconducting TES film has been thermally isolated from the surroundings by mounting it %%@
on a thin dielectric membrane, usually made of amorphous silicon nitride (SiN$_\mathrm{x}$) due to ease of fabrication. This %%@
membrane limits the thermal conductance to the bath, and  
 is therefore critical for the operation of the devices.

The phonon transport in the membrane can be either diffusive or ballistic, depending on how easily phonons are scattered in the %%@
sample. At the low temperatures where TESes are operated ($\sim 0.1 $ K), the bulk scattering mechanisms (mass impurities, %%@
phonon-phonon scattering, phonon-two-level system scattering)  become very weak, \cite{ber} leading to surface limited thermal %%@
conduction. If, in addition, the surfaces are smooth on the length scale of the dominant thermal phonons, the surface scattering %%@
is mostly specular, and phonon transport becomes ballistic.  \cite{zim} Ballistic phonon transport has been observed for %%@
crystalline bulk samples a long time ago \cite{vongut}, but was also recently shown to be valid for some SiN$_\mathrm{x}$ %%@
membranes  \cite{gild,sron}. For thermal sources such as TES film radiators in the ballistic limit, the  analogy with photon %%@
thermal black-body radiation is apparent if the substrate is three-dimensional. Then the emitted power has the typical %%@
Stefan-Bolzmann form $P=A\sigma T^4$, where for phonons $\sigma=\pi^5k_B^4/(15h^3)\Sigma e_i/c_i^2$, summing over the different %%@
phonon modes with speeds of sound $c_i$ and radiator emissivities $e_i$. \cite{wolfe}  
On the other hand, it is less clear what happens to ballistic phonon transport when the substrate becomes two-dimensional. 
This is a realistic concern, as the dominant emitted phonon wavelength is of the order of $1 \mu$m for SiN at 100 mK, \cite{note} %%@
comparable with a typical membrane thickness, so that many practical devices are at least approaching the 2D limit.    

In this paper, we describe a theory for operating TES detectors on thin membranes, spanning the transition from a fully 3D to a %%@
fully 2D substrate (membrane), using elasticity theory. The eigenmodes of a thin membrane are no longer the usual plane wave %%@
phonons, but are so called Lamb modes, with non-trivial displacement fields and dispersion relations. \cite{Auld} For this reason, %%@
thermal conduction \cite{kmnima} and the phononic heat capacity \cite{fefe} are strongly affected, leading to the interesting %%@
effect that radiated power and thermal conductance have an absolute minimum as a function of the membrane thickness.     
Because of this effect, the noise equivalent power (NEP) also has a minimum, leading to the notion of an optimal membrane %%@
thickness for TES bolometers. However, the effective time constant also has a maximum at the same point, thus, one needs to make %%@
sure that the negative effect of slowing the detector down is not critical for the application. We also calculated the influence %%@
of thin membranes on the phonon noise limited energy resolution of TES calorimeters, and found that the effect is weak, especially %%@
around the optimal detector temperature.

%\section{The phonon modes of a membrane}

In isotropic 3D bulk systems there are three independent phonon modes, two transversally
and one longitudinally polarized, with sound velocities $c_t$ and $c_l$, respectively. 
In the presence of boundaries, the bulk phonon modes
couple to each other and form a new set of eigenmodes, which in the case of a free standing
membrane are horizontal shear modes ($h$) and symmetric ($s$) and antisymmetric
($a$) Lamb modes \cite{Auld}.
The frequency $\omega$ for the $h$ modes is simply 
$\omega= c_t\sqrt{\kpar^2+(m\pi/d)^2}$,
where $\kpar$ is wave vector component parallel to membrane surfaces,
$d$ is the membrane thickness and the integer $m$ is the branch number. However, the dispersion
relations of the $s$ and $a$ Lamb modes cannot be given in a closed analytical form, but have to be 
calculated numerically. \cite{PhysRevB} The lowest three branches, dominant for thin membranes at low temperatures, have low %%@
frequency analytical expressions:
%
%\begin{eqnarray}
%\frac{\tan(k_td/2)}{\tan(k_ld/2)}
% &=& -\frac{4k_tk_l\kpar^2}{(k_t^2-\kpar^2)^2}\label{eqn_disp_s}
%\end{eqnarray}
%
%for the symmetric Lamb modes and
%
%\begin{eqnarray}
%\frac{\tan(k_td/2)}{\tan(k_ld/2)}
% &=& -\frac{(k_t^2-\kpar^2)^2}{4k_tk_l\kpar^2}\label{eqn_disp_a}
%\end{eqnarray}
%
%for the antisymmetric Lamb modes. Here $k_t$ and $k_l$ are the perpendicular part
%of the wave vectors of the transversal and longitudinal partial waves, respectively. They
%are connected by the Snell's law of accoustics, $c_t^2(k_t^2+\kpar^2)=c_l^2(k_l^2+\kpar^2)$.

%In the low energy limit we can give analytic expressions for the respective lowest
%branch of the dispersion relation of each of the three membrane modes. These are
%\cite{PhysRevB.70.125425}
%
\begin{subequations}\label{eqn_low_energy_disp}
\begin{eqnarray}
\omega_{h,0}
 &=& c_t\kpar\label{eqn_low_energy_disp_h}\\
\omega_{s,0}
 &=& c_s\kpar\label{eqn_low_energy_disp_s}\\
\omega_{a,0}
 &=& \frac{\hbar}{2m^\star}\kpar^2\,\label{eqn_low_energy_disp_a}
\end{eqnarray}
\end{subequations}
where $c_s=2c_t\sqrt{(c_l^2-c_t^2)/c_l^2}$ is the effective sound velocity of the $s$ mode, and
$m^\star=\hbar\left[2c_td\sqrt{(c_l^2-c_t^2)/3c_l^2}\right]^{-1}$ is an effective mass for the $a$-mode "particle". 
This lowest $a$-mode with its quadratic dispersion is mostly responsible for the non-trivial behavior for the detector performance %%@
described below.  
%
%\section{Radiative power flow in the membrane}
%
%In Ref.\ [\onlinecite{KuehnMaasilta.NuclInstMeth}] we studied, how an electronic nanoscale 
%device (a detector) is cooled by the membrane phonons. For this we assumed ideal
%electron-phonon coupling (i.e. the phonon gas directly below the detector has the
%same temperature as the electron gas of the detector) and radiative heat transfer in
%the membrane \cite{ApplPhysLett86.251903.Hoevers}. 

\begin{figure}
\begin{center}
%The bounding box should be 230x130
\resizebox{90mm}{!}{\includegraphics{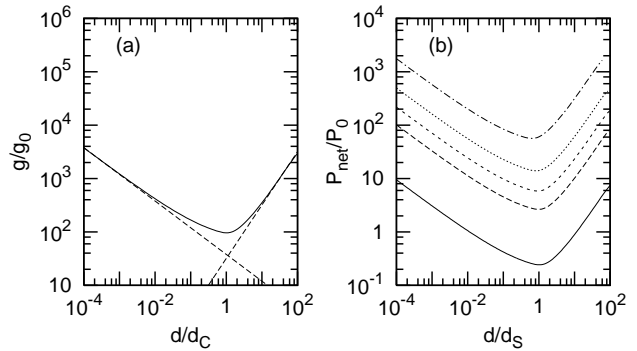}}
\end{center}
\caption{(a) The differential thermal conductance $g$ of a ballistic radiator as function of the membrane thickness $d$.
The crossover thickness $d_C=\hbar c_t/2k_BT$. $g$ is normalized with $g_0=(k_Blc_t/8\pi^2)(k_BT/\hbar c_t)^2$. The asymptotic 2D %%@
and 3D limits, computed from Eqs. \ref{eqn_P_2D} and \ref{eqn_P_3D} are indicated by the dashed lines.
(b) The net emitted power $P_{net}$ of a detector at temperature $T_D$ connected to the
substrate at temperature $T_S$ as function of membrane thickness $d$, for temperature ratios
$T_D/TS=1.01$, $1.1$, $1.2$, $1.4$ and $2.0$. Curves with higher $T_D/T_S$
lie above curves with lower $T_D/T_S$.  $d_S=\hbar c_t/2k_BT_S$ and 
$P_0=l(k_BT_S)^3/(2\pi\hbar^2 c_t)$.}\label{fig_1}
\end{figure}

To simplify the discussion, we assume that the thermal conductance is only limited by the membrane itself, in other words we do %%@
not consider the effects of electron-phonon non-equilibrium or thermal gradients within the TES film \cite{jenni}, or of boundary %%@
resistance between the TES film and the membrane, and the membrane and the supporting 3D substrate. The importance of these added %%@
effects depends critically on the materials and detector geometry, and can in principle be minimized. With these assumptions   
the total heat flow out of the detector is

\begin{equation}
P=\frac{l}{2\pi^2}\sum_{\sigma,m}\lint{0}{\infty}{k_\parallel} k_\parallel
     \hbar\omega_{\sigma,m}\left|\frac{\partial\omega_{\sigma,m}}{\partial k_\parallel}\right|n(\omega,T),
     \label{eqn_P_of_T}
\end{equation}
where $l$ is the circumference of the detector, $n(\omega,T)$ is the Bose-Einstein distribution and $\sigma$ and $m$ are the mode %%@
and branch indices. \cite{note2} This expression can be used to compute the transition from 3D to 2D if enough branches are used. 
If the membrane is thin and temperature low, i.e. $Td\ll\hbar c_t/2k_B$, only
the lowest branches (Eqs.\ \ref{eqn_low_energy_disp}) are occupied, and we are fully in the 2D limit, in which case 

\begin{eqnarray}
P_\mathrm{2D} 
 &=&  \frac{l\hbar}{2\pi^2}\left[\!\left(\frac{1}{c_t}+\frac{1}{c_s}\right)\!
      \Gamma(3)\zeta(3)\left(\frac{k_BT}{\hbar}\right)^3\right.\nonumber\\
     &&\left.+\sqrt{\frac{2m^*}{\hbar}}\Gamma\!\left(\frac{5}{2}\right)\!
      \zeta\!\left(\frac{5}{2}\right)
      \!\left(\frac{k_BT}{\hbar}\right)^{\!5/2}\right]\,.\label{eqn_P_2D}
\end{eqnarray}
Note that the effective mass of the lowest $a$ mode depends on the membrane thickness and hence
in the 2D limit $P\propto 1/\sqrt{d}$. In the 3D limit $Td\gg\hbar c_t/2k_B$,
the dominant phonon wavelength is much smaller than $d$, leading to decoupling of  the longitudinal and 
transversal modes and
\begin{eqnarray}
P_\mathrm{3D}
 &=& \frac{\pi^2ld\hbar}{120}\!\left(\frac{2}{c_t^2}+\frac{1}{c_l^2}\right)
     \!\left(\frac{k_BT}{\hbar}\right)^4\,.\label{eqn_P_3D}
\end{eqnarray}
As expected, $P_\mathrm{3D}\propto d$. This means that decreasing the membrane thickness at
a fixed temperature, the radiated power will first decrease with $d$, then reach a global minimum
and will increase again, if we decrease $d$ further. The minimum is approximately at the
2D-3D crossover thickness $d_C\equiv \hbar c_t/(2k_B T)$. The same behavior occurs for the differential thermal conductance
$g=dP/dT$.
In the 2D-3D crossover range, we have computed
$P$ and $g$  numerically, using the lowest 100 branches.  Fig.\ \ref{fig_1}(a), we plot
$g$ as a function of $d$, showing an increase of $g$ by $\sim$ factor of 5 when decreasing $d$ from $d_C$ to $10^{-2} d_C$, %%@
corresponding to changing $d$ from 240 nm to 2.4 nm at 100 mK for SiN$_\mathrm{x}$.

The net phonon cooling power of the detector is the difference between the power flow out
of the detector and the power flow into the detector that is emitted from the substrate.
Hence, if we denote
the detector temperature by $T_D$ and the substrate temperature by $T_S$, the total cooling
power is $P_\mathrm{net}=P(T_D)-P(T_S)$. In Fig.\ \ref{fig_1}(b) we plot $P_\mathrm{net}$
as function of $d$ for different ratios $T_D/T_S$. As we have two different temperatures, we
chose to normalize the membrane thickness with respect to $T_S$ and define
the quantity $d_S\equiv\hbar c_t/2k_B T_S$. The thickness dependence is the same as for $g$, and as one would expect,
the radiated power increases with increasing $T_D$. The place of the
minimum is slightly shifted from curve to curve, as the 2D-3D crossover is dominated by the detector phonons at high $T_D/T_S$. 

%\section{Detector parameters}

A useful way to operate TES detectors is by voltage biasing. This leads to negative electrothermal feedback (ETF), which speeds up %%@
the detectors and improves their sensitivity and energy resolution \cite{enss}. For TES on thin ballistic membranes, the basic %%@
operational theory applies, as long as one takes into account the correct formulas for $P_{net}$ and $g$ as discussed above. %%@
Especially interesting is the question: Will the non-trivial thickness dependence lead to observable effects for the important %%@
detector parameters such as the noise equivalent power NEP, effective time constant $\tau_{\mathrm{eff}}$ and energy resolution %%@
$\Delta E$? In the following we discuss this issue, noting that we only consider here the contributions from the thermodynamic %%@
noise generated by the phonon thermal transport in the membrane, the so called phonon noise. Other noise mechanisms can be easily %%@
added to the discussion using known formulas \cite{enss}. 

\begin{figure}[h]
\begin{center}
%The bounding box should be 230x130
\resizebox{90mm}{!}{\includegraphics{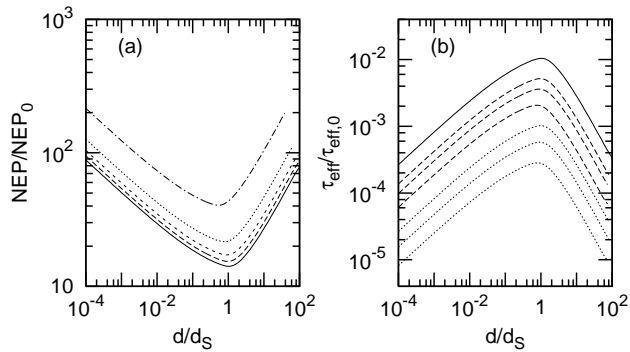}}
\end{center}
\caption{
(a)The ballistic phonon NEP of a TES as function of $d$ for temperature ratios
$T_D/T_S=1.0$, $1.1$, $1.2$, $1.4$ and $2.0$, ordered from bottom to top. The normalization constant is
$\mathrm{NEP}_0=\sqrt{l/c_t}(k_BT_S)^2/(2\pi\hbar)$.
(b) The effective time constant $\tau_\mathrm{eff}$ as function of $d$
for the same ratios $T_D/T_S=1$ (solid line) and $1.1$, $1.2$ and $1.5$ for 
$\alpha=10$ (dashed lines) and $\alpha=100$ (dotted lines). Curves with higher
$T_D/T_S$ below curves with lower $T_D/T_S$ for the same value of $\alpha$.
 $\tau_{\mathrm{eff},0}=8\pi^2\hbar^2c_t\gamma V_\mathrm{el}/(lk_B^3T_S)$.
}\label{fig_2}
\end{figure}
The phonon noise power spectral density for ballistic transport
can be calculated from microscopic considerations, \cite{boyle,maa} leading to an equation for the phonon NEP:
\begin{eqnarray}
\mathrm{NEP}
 &=& \sqrt{2k_B[T_D^2g(T_D)+T_S^2g(T_S)]}\,.
 \label{NEPeq}
\end{eqnarray}
In Fig.\ \ref{fig_2}(a), we plot the NEP as a function of $d$. It has a clear minimum, just like $g$, as expected from the %%@
simplicity of Eq. (\ref{NEPeq}). In the 3D limit the NEP becomes
proportional to $\sqrt{d}$, whereas in the 2D limit it 
approaches $d^{-1/4}$. The absolute minimum in the NEP means that it is possible to define the optimal membrane thickness %%@
$d_{opt}$, which gives the highest sensitivity for a TES bolometer at fixed $T_D$ and $T_S$. For an equilibrium bolometer %%@
($T_D=T_S$) $d_{opt} \approx d_S$, whereas for $T_D > T_S$, $d_{opt} \approx \hbar c_t/(2k_BT_D)$. For SiN$_x$ membranes and %%@
$T_D=T_S=0.1 K$, $d_{opt}= 240$ nm. 

In addition to sensitivity, an important characteristic of a bolometer is its response time $\tau_{\mathrm{eff}}$. For a voltage %%@
biased TES of volume $V$ with its electronic heat capacity $C_{el}=\gamma VT_D$ being dominant, in the perfect voltage bias and %%@
low inductance limit \cite{enss} it is given by      
$\tau_{\mathrm{eff}}= C_{el}/(g(T_D)+\alpha P_{net}/T_D)$, where $\alpha=T/R dR/dT$ is the steepness parameter of the transition %%@
under bias. Fig.\ \ref{fig_2}(b) presents $\tau_\mathrm{eff}$ as function of $d$ for several values of $T_D/T_S$ and two examples %%@
of $\alpha=10,100$.  Now 
$\tau_\mathrm{eff}$ has a maximum around $\hbar c_t/(2k_BT_D)$ and is proportional to $\sqrt{d}$ in the
2D limit and to $1/d$ in the 3D limit. Thus, if one wants to operate near $d_{opt}$, limitations of the time constant need to be %%@
considered. Fortunately, biasing strongly into the ETF (high $\alpha$), helps to reduce $\tau_\mathrm{eff}$ significantly.

\begin{figure}[h]
\begin{center}
%The bounding box should be 230x130
\resizebox{90mm}{!}{\includegraphics{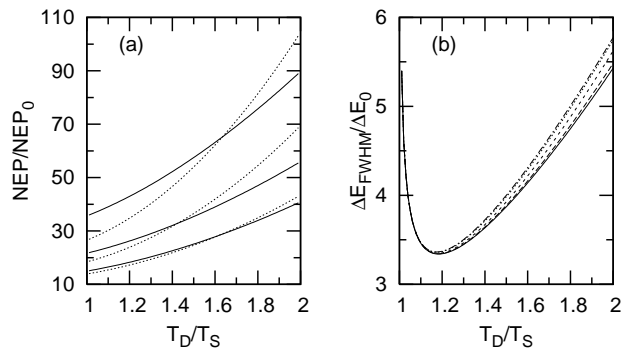}}
\end{center}
\caption{(a) The phonon NEP vs. $T_D/T_S$ for membrane
thicknesses $d/d_S=0.004$, $0.04$, $0.4$, $1$, $4$ and $40$. The curves for
$d/d_S<1$ are plotted with solid lines with higher curves corresponding to lower $d$, while 
dotted lines represent curves with $d/d_S\ge1$ with lower curves corresponding to lower $d$.
(b) The high ${\cal L}$ FWHM energy resolution $\Delta E_{FWHM}$ of a TES as function of the ratio
$T_D/T_S$ for the membrane thicknesses 
$d/d_S=0.04$, $0.4$, $1$, $4$ and $40$, ordered from the bottom to the top. $\Delta E_0=2.8
\sqrt{\gamma V k_B T_S^3/\alpha_I}$.}\label{fig_3}
\end{figure}

In Fig.\ \ref{fig_3}(a), we show the NEP as function of $T_D$  for different values of $d$. In general
the NEP increases with $T_D$. 
For a fixed value of $T_D/T_S$ we can always find two membrane thicknesses that
yield the same NEP, one with $d<d_S$ and one with $d>d_S$. However, the $T_D$ dependence of these two values are quite different, %%@
as can be seen by comparing the slopes of the curves at crossing points.

Finally, the phonon noise limited full width half maximum (FWHM) energy resolution of an optimally filtered TES calorimeter %%@
\cite{enss} for high loop gain ${\cal L}=\alpha P_{net}/(g(T_D)T_D) \gg 1$  is
\begin{equation}
\Delta E_\mathrm{FWHM}= 2.355\sqrt{k_BT_D^2C_{el}}\frac{2}{\sqrt{\alpha_I}}
     \left(\frac{\mathrm{NEP}^2}{4k_BT_DP_\mathrm{net}}\right)^{1/4} \hspace{-4mm},
\end{equation}

where $\alpha_I=T/R \partial R/\partial T$. 
 $\Delta E_\mathrm{FWHM}$ is plotted as a function of $T_D$ in
Fig.\ \ref{fig_3}(b) for different membrane thicknesses. It has a minimum just below 
$T_D/T_S\approx 1.2$, almost independent of $d$ and in agreement with the 3D ballistic result \cite{ilariNIMA}, so that an optimal %%@
$T_D$ for a fixed $T_S$ can be defined. For higher ratios $T_D/T_S$, the thinner membranes decrease $\Delta E_\mathrm{FWHM}$ %%@
slightly, by about 6 \% at $T_D/T_S=2$ from $d=40 d_S$ to $d=0.04 d_S$.    

Discussions with D.-V. Anghel and M. Manninen are acknowledged. This work was supported by ESA ESTEC contract no. 16759/02/NL/PA, %%@
the Academy of Finland project No. 118665 and EU project No. 505457-1.


\begin{thebibliography}{00}
\bibitem{note} The transversal and longitudinal sound velocities for low-stress SiN$_\mathrm{x}$ are 
$c_t=6200$ m/s and $c_l=10300$ m/s, respectively \cite{wenzel}, giving dominant wavelengths $\lambda=hc_i/(2.82k_BT)$ as 
$\lambda_t=1.0 \mu$m and $\lambda_l=1.8 \mu$m at 0.1 K.
\bibitem{note2} Note that the prefactor is $\frac{l}{2\pi^2}$ instead of the factor $\frac{l}{8\pi}$
that was mistakenly given in Ref.\ [\onlinecite{kmnima}].
\end{thebibliography}
\end{document}